\documentclass[preprint,aps]{revtex4}

\usepackage{graphicx}

\begin{document}

\title{Electronic Structure and Nematic Phase Transition in Superconducting Multiple-Layer FeSe Films Grown by Pulsed Laser Deposition Method}

\author{Bing Shen$^{1,2,\dag}$,  Zhongpei Feng$^{1,2,\dag}$, Jianwei Huang$^{1,2}$, Yong Hu$^{1,2}$, Qiang Gao$^{1,2}$, Cong Li$^{1,2}$, Yu Xu$^{1,2}$, Guodong Liu$^{1}$, Li Yu$^{1}$,  Lin Zhao$^{1,*}$,  Kui Jin$^{1,*}$ and X. J. Zhou$^{1,2,3,*}$
}

\affiliation{
\\$^{1}$National Lab for Superconductivity, Beijing National Laboratory for Condensed Matter Physics, Institute of Physics,
Chinese Academy of Sciences, Beijing 100190, China
\\$^{2}$University of Chinese Academy of Sciences, Beijing 100049, China
\\$^{3}$Collaborative Innovation Center of Quantum Matter, Beijing100871, China
\\$^{\dag}$These people contribute equally to the present work.
\\$^{*}$Corresponding authors: LZhao@iphy.ac.cn, kuijin@iphy.ac.cn,  XJZhou@aphy.iphy.ac.cn.
}

\date{\today}

\begin{abstract}
We report comprehensive angle-resolved photoemission investigations on the electronic structure of single crystal multiple-layer FeSe films grown on CaF$_2$ substrate by pulsed laser deposition (PLD) method.  Measurements on FeSe/CaF$_2$ samples with different superconducting transition temperatures T$_c$ of 4 K, 9 K and 14 K reveal electronic difference in their Fermi surface and band structure.   Indication of the nematic phase transition is observed from temperature-dependent measurements of these samples; the nematic transition temperature is 140$\sim$160 K, much higher than $\sim$90 K for the bulk FeSe.  Potassium deposition is applied onto the surface of these samples; the nematic phase is suppressed by potassium deposition which introduces electrons to these FeSe films and causes a pronounced electronic structure change. We compared and discussed the electronic structure and superconductivity of the FeSe/CaF$_2$ films by PLD method with the FeSe/SrTiO$_3$ films by molecular beam epitaxy (MBE) method and bulk FeSe.  The PLD-grown multilayer FeSe/CaF$_2$ is more hole-doped than that in MBE-grown multiple-layer FeSe films.  Our results on FeSe/CaF$_2$ films by PLD method establish a link between bulk FeSe single crystal and FeSe/SrTiO$_3$ films by MBE method, and provide important information to understand superconductivity in FeSe-related systems.
\end{abstract}


\maketitle

\newpage

The iron chalcogenides (FeSe-based) are one of two major classes of the iron-based superconductors in addition to the iron pnictides (FeAs-based)\cite{Hsu2008,Mazin2009,Guo2010,Mizuguchi2010,Mazin2010,Song2011,Paglione2010,Wang2012}. Recently great efforts have been focused on the iron chalcogenides, including FeSe single crystals,  FeSe films, and FeSe-related materials like A$_x$Fe$_{2-y}$Se$_2$ and (Li,Fe)OHFeSe\cite{Paglione2010,Wang2011,Ying2012,Dagotto2013,Dong2014,Lu2015}. Bulk FeSe has a superconducting transition temperature of $\sim$8 K\cite{Hsu2008} at ambient pressure; its T$_c$ can be significantly enhanced up to 38 K under high pressure\cite{Medvedev2009,Okabe2010} or up to 48 K by the gating method\cite{Lei2016}. Single-layer FeSe films grown on SrTiO$_3$ substrate by molecular beam epitaxy (MBE) method show signatures of superconductivity up to 65 K\cite{Wang2012,Liu2012,He2013,Tan2013}. Multiple-layer FeSe/SrTiO$_3$ films are thought to approach bulk materials when they are thick enough, however, FeSe/SrTiO$_3$ films made by MBE method are not superconducting when their thickness is two or more layers\cite{Wang2012}, even above 50 layers\cite{Tan2013}. Study on the origin of the similarity and difference between the bulk FeSe and FeSe/SrTiO$_3$ films by MBE method may provide important information to understand their underlying superconductivity mechanism.

Recently, pulsed laser deposition (PLD) method was developed to grow FeSe films on CaF$_2$ substrate\cite{Jin2017}. Their superconductivity can be tuned by slightly varying the preparation conditions,  resulting in FeSe/CaF$_2$ films from non-superconducting to superconducting with different T$_c$s, even up to 15 K that is obviously higher than that of bulk FeSe. XRD measurements show that the lattice constant of $c$ axis becomes larger for those samples with higher T$_c$s. It is interesting to investigate the electronic structure of these PLD-grown FeSe/CaF$_2$ films,  and compare with that of bulk FeSe and MBE-grown FeSe/SrTiO$_3$ films in order to understand the difference of superconductivity in these samples.

In this paper, we report high resolution angle-resolved photoemission (ARPES) measurements on multiple-layer FeSe films grown on CaF$_2$ substrate by PLD method  with different superconducting critical temperature T$_c$ at 4 K, 9 K and 14 K\cite{Jin2017}.   Fermi surface and band structure along representative momentum cuts are measured for FeSe/CaF$_2$ samples with different T$_c$s. They are qualitatively similar to the electronic structure of bulk FeSe\cite{Maletz2014,Nakayama2014,Shimojima2014,Watson2015,Zhang2015,Watson2016} and multi-layer MBE-grown FeSe/SrTiO$_3$ films\cite{Tan2013,LiuX2014,Peng2014,Miyata2015}.  Indication of the nematic phase transition is observed from temperature-dependent measurements of band-splitting in these samples; the nematic transition temperature is $\sim$150 K, much higher than $\sim$90 K for bulk FeSe.  By depositing potassium onto the surface of these samples, we find that the nematic phase is suppressed by potassium deposition which introduces electrons into these FeSe films and causes a pronounced electronic structure change.   We will compare and discuss on the  electronic structure and superconductivity of the FeSe/CaF$_2$ films by PLD method with the FeSe/SrTiO$_3$ films by MBE method and bulk FeSe.  Our results on FeSe/CaF$_2$ films by PLD method establish a link between bulk FeSe single crystal and FeSe/SrTiO$_3$ films by MBE method, and provide important information to understand superconductivity in FeSe-related systems.

The multiple-layer FeSe films were grown on CaF$_2$ substrate with a typical size of $5 mm\times5 mm$ using PLD method reported in a separate paper\cite{Jin2017}.  The thickness of these FeSe/CaF$_2$ films is about 160 nm that contains about 300 FeSe layers. The FeSe films used in our experiments were characterized by resistivity measurements; three kinds of films were measured with a  T$_c$ at 4 K (Fig. 1a), 9 K (Fig. 2a)  and 14 K (Fig. 3a). The angle-resolved photoemission measurements were carried out by using our lab system equipped with a Scienta R4000 electron energy analyzer\cite{Liu2008}. We used HeI resonance line as the light source which gives a photon energy of h$\nu$=21.218 eV. The light is partially polarized with the electric field vector mainly in the plane of the sample surface. The energy resolution was set at 10 meV and the angular resolution was $\sim$0.3$^\circ$. The Fermi level was referenced by measuring the Fermi edge of a clean polycrystalline gold that is electrically connected to the sample. The FeSe/CaF$_2$  film samples were prepared by cleaving  \emph{in situ} at 20 K and measured in vacuum at different temperatures with a base pressure better than $6\times10^{-11}$ Torr.


Figure 1 shows the measured electronic structure of the PLD-grown FeSe/CaF$_2$ film with a T$_c$=4 K (denoted as S4K sample hereafter). Fig. 1b1 and 1b2 show Fermi surface mapping measured at 30 K and 210 K, respectively.  The Fermi surface is obtained by integrating the spectral wight within [-20meV, 10meV] energy window with respect to the Fermi level.  The band structure measured at different temperatures,  along three momentum cuts crossing $\Gamma$,  M2 and M3 are presented in Fig. 1c, Fig. 1d and Fig. 1e, respectively. The location of the three momentum cuts are marked in Fig. 1b1 by the red lines. Three hole-like bands are resolved near the $\Gamma$ point (Fig. 1c) which are denoted as $\alpha$, $\beta$ and $\gamma$ in Fig. 1c1. While the $\alpha$ and $\beta$ bands which are considered as dominant $d_{xz}/d_{yz}$ orbital character\cite{Watson2015,MYi2015,Liu2015} cross the Fermi level, the $\gamma$ band which has dominant $d_{xy}$ orbital character\cite{Watson2015,MYi2015,Liu2015}is flat and lies $\sim$45 meV below the Fermi level.  These bands give rise to hole-like Fermi pockets in the Fermi surface mapping (Fig. 1b).    The observation of three hole-like bands around  $\Gamma$  near the Fermi level is qualitatively consistent with the theoretical calculation\cite{Subedi2008} and  bulk FeSe results\cite{Maletz2014,Nakayama2014,Shimojima2014,Watson2015,Zhang2015,Watson2016},  and also similar to other iron-based compounds like BaFe$_2$As$_2$\cite{Liu2009}.   Three bands can be clearly resolved in the measured results near M2 (Fig. 1d) and M3 (Fig. 1e) at low temperature, labeled as $\varepsilon$, $\delta$ and $\eta$ in Fig. 1d1. These bands are similar at M2 and M3, but have different spectral weight due to photoemission matrix element effect\cite{Damascelli2003}. $\delta$ band is hole-like that crosses the Fermi level while the other hole-like band $\eta$  lies $\sim$60 meV below the Fermi level. These two bands are considered to come from band splitting of bands with $d_{xz}/d_{yz}$ orbital character in the tetragonal phase\cite{Shimojima2014,Watson2015,MYi2015,Liu2015}. In this case, the band with $d_{xy}$ orbital character (dashed line in Fig. 1d1 and 1e1) is not resolved which can be due to matrix element effect\cite{Damascelli2003} or orbital selective Mott physics\cite{MYi2013,RYu2013}. At present, there is a debate on the band splitting and band assignment at M point\cite{Fedorov2016}; further efforts are needed for resolving these issues.  In addition to these two bands, the electron-like band $\varepsilon$ (which is considered to have $d_{xz}/d_{yz}$ orbital character\cite{Shimojima2014,Watson2015,MYi2015,Liu2015}) is observed with its bottom $\sim$13 meV below the Fermi level.  At low temperature, these band structures around M2 and M3 form Fermi surface around the Brillouin zone corners: one small electron-like pocket centering at the M point and four intensity spots around the M point. The observation of such Fermi surface and band structure is not consistent with band structure calculation results in paramagnetic state\cite{Subedi2008}, but are similar to those  from bulk FeSe in the nematic phase\cite{Maletz2014,Nakayama2014,Shimojima2014,Watson2015,Zhang2015,Watson2016}, and  multiple-layer MBE-grown FeSe/SrTiO$_3$ films at low temperature\cite{Tan2013,Peng2014,Zhang2016}.

The observed well-defined Fermi surface and band structure demonstrate that the PLD-grown FeSe/CaF$_2$ films we have measured are single crystal-like, which is highly oriented both  along \emph{c} direction and within the \emph{ab} plane that are well suited for ARPES measurements.   In MBE-grown FeSe/SrTiO$_3$ films,  there are two distinct electronic structures that have been identified: one is labeled as N phase that is characterized by ``four intensity spots"  feature at M points when the doping level is low while the other is called S phase that is characterized by an electronic pocket around M when sufficient electron doping is introduced\cite{He2013,LiuX2014}.  The N phase is commonly observed in bulk FeSe\cite{Maletz2014,Nakayama2014,Shimojima2014,Watson2015,Zhang2015,Watson2016}, MBE-grown multiple-layer FeSe/SrTiO$_3$ films\cite{Tan2013,Miyata2015,Zhang2016} and actually in many parent compounds of the iron-based superconductcors\cite{LiuDF2016}. Overall, the electronic structure of our PLD-grown FerSe/CaF$_2$ films is similar to that of bulk FeSe and MBE-grown multiple-layer FeSe/SrTiO$_3$ films, i.e., their electronic structures resemble that of the N phase.

Figure 2 and figure 3 show Fermi surface and temperature-dependent evolution of the band structure along three momentum cuts for the PLD-grown FeSe/CaF$_2$ films with T$_c$=9 K (denoted as S9K) and T$_c$=14 K (denoted as S14K), respectively. Overall speaking, these PLD-grown FeSe films with different T$_c$s show qualitatively similar electronic structure and the temperature dependence to that of S4K shown in Fig. 1. In order to make a quantitative comparison to identify possible T$_c$-dependent difference, Fig. 4 directly compares the band structure along three momentum cuts for the S4K, S9K and S14K samples measured at a low temperature 30 K and a high temperature 150 K.  Although the overall band structure is similar for the three samples, there are some notable differences: (1). For the bands near $\Gamma$ point, little change is observed for the outer $\beta$ band that is a dominant contributor of the hole-doping (Fig. 4a1, 4a4 and 4a7). The flat band $\gamma$ appears to shift upwards with a band top moving from $\sim$47 meV binding energy for S4K sample (Fig. 4a1)  to $\sim$36 meV for S14K sample (Fig. 4a7). (2). For the  $\varepsilon$ electron band seen near M2 and M3,  overall the band is very shallow which contributes small amount of electrons.  There seems to be a tendency that this band bottom shifts upwards to the Fermi level from S4K sample (Fig. 4a2 and 4a3), to S9K (Fig. 4a5 and 4a6) and S14K (Fig. 4a8 and 4a9) samples, implying that electrons decrease with increasing T$_c$. We note that there may be other electron-like band(s) that are not resolved in our measurements;  (3). The crossing point of the $\delta$ band at the Fermi level appears to get closer to the M point from S4K (Fig. 4a3), to S9K (Fig. 4a6) to S14K (Fig. 4a9) samples.  Since this band is related to the tiny strong intensity spots that form Dirac cones, it has weak effect on the charge carrier.  (4). The three samples show slight difference on the temperature dependence of bands at M2 and M3 points, as we will discuss below.

Detailed temperature-dependent measurements of the bands along the three momentum cuts (Fig. 1) indicate that,  while the three hole-like bands near $\Gamma$ do not show obvious change with temperature (Fig. 1c1-c8 from 30 K to 170 K), the bands near M2 and M3 points show dramatic temperature dependence that is related to the nematic phase transition. This is characterized by the band splitting between the $\delta$ and $\eta$ bands below the nematic transition temperature. For the band at M3 (Fig. 1e), the $\delta$ and $\eta$ bands are clear.  At a low temperature of 30 K,  there is a band splitting between $\delta$ and $\eta$ bands with a distance of $\sim$50 meV between the top of  $\delta$ band and the top of $\eta$ band.  With increasing temperature, the band splitting gets smaller and the $\delta$ and $\eta$ bands nearly merge to each other at around 150 K-170 K that gives a nematic transition temperature around 160 K.  By the same criterion,  the nematic transition temperature determined from the $\delta$ and $\eta$ band splitting is about 150 K for the S9K sample (Fig. 2e), and $\sim$140 K for the S14K sample (Fig. 3e).  The nematic transition temperature can also be estimated from the splitting between the $\varepsilon$ electron band and the $\eta$ hole band that are clear in the measurements at M2 points (Fig. 1d).  For the S4K sample, at a low temperature of 30 K, these two bands are well separated with a distance of $\sim$50 meV between the $\varepsilon$ band bottom and the $\eta$ band top (Fig. 1d1).  At a high temperature like 170 K (Fig. 1d8),  these two bands merge to form a single band.  In between, the two bands start to merge at a temperature between 150 K and 170 K (Fig. 1d1-d8) so the nematic transition temperature is estimated to be about 160 K.  In the same manner, the nematic transition temperature determined for the S9K (Fig. 2d1-d7) and S14K (Fig. 3d1-d10) can be estimated to be $\sim$150 K and $\sim$140 K, respectively. The nematic transition temperature determined this way is consistent with that from the above $\delta$ and $\eta$ band splitting.  We note that, because the transition is not abrupt and the temperature coverage is not dense, there is a large error bar  in determining the value of the nematic transition temperature  with an uncertainty of $\pm$10 K or even slightly higher. But the overall trend of the nematic transition temperature decrease from S4K, to S9K to S14K seems to be more robust, as comparing the bands measured at M2 at 150 K (Fig. 4b2, b5 and b8).  The  $\varepsilon$ and $\eta$  bands barely merge for S4K (Fig. 4b2),  but fully merge  for the S14K sample (Fig. 4b8), suggesting that the nematic transition temperature of the S4K sample is higher than that of the S14K sample.

In order to quantitatively keep track on the temperature evolution of the band splitting and the nematic phase transition, Fig.  5 shows temperature dependence of photoemission spectra (energy distribution curves, EDCs) at a momentum k$_1$ near the M3 point. The location of the k$_1$ point is marked in Fig. 5c where the band that crosses the M3 point of the S14K sample is re-plotted.  In Fig. 5c,  k$_0$ represents the M point and  $\delta$ band, $\eta$ band and $\varepsilon$ band are clearly observed with  the dashed lines as the guide to the eyes for the $\delta$ and $\eta$ bands.  As seen in Fig. 5c,   the EDC at the momentum k$_1$  covers both the $\delta$ band and $\eta$ band. Fig. 5a and 5b show EDCs at the k$_1$ point measured at different temperatures for the S4K and S14K samples, respectively.  At low temperatures, the $\delta$ and $\eta$ bands show up in EDCs as two peaks; its distance can be used as a measure of the band splitting between the $\delta$ and $\eta$ bands.  With increasing temperature, these two peaks get closer and eventually merge to each other at a temperature that correspond to the nematic transition temperature.   Quantitative results of the band splitting at k$_1$ point for S4K, S9K and S14K samples are summarized in Fig. 5d.  The nematic transition temperature for S4K, S9K and S14K samples are $\sim$160 K, $\sim$150 K and $\sim$140 K, respectively.  In order to facilitate comparison among different compounds, the band splitting is best  determined at the high-symmetry M point\cite{LiuDF2016}. To do this, the band top of the $\delta$ band is extrapolated by fitting the observable band with a parabola, as shown in Fig. 5c. The band splitting size at M point (k$_0$) for the three samples is also plotted in Fig. 5d.  At 30 K, the band splitting at M gradually decreases from $\sim$53 meV for S4K, to $\sim$48 meV for S9K, and to $\sim$43 meV for S14K samples.   Our results indicate that the nematic transition temperature in PLD-grown FeSe/CaF$_2$ films (140$\sim$160 K) is enhanced compared with that of bulk FeSe (90 K).  Also with the increasing of T$_c$, the nematic transition in FeSe/CaF$_2$ films appears to be suppressed.

It has been shown that introduction of electrons into bulk FeSe by gating method can enhance its T$_c$ from original 8 K to 48 K, and induce superconductor-insulator transition\cite{Lei2016}.  For the MBE-grown FeSe/SrTiO$_3$ films, it was also found that introduction of electrons by vacuum annealing\cite{He2013} or potassium deposition\cite{Miyata2015,Tang2015,Song2016,Wen2016,ZhangWH2016} can transform non-superconducting films into superconducting ones, and can further enhance their T$_c$s.  Therefore, we investigated the doping evolution of the electronic structure for the PLD-grown FeSe/CaF$_2$ films by depositing potassium on their surfaces.  Fig. 6 shows the Fermi surface and band structure for the S4K sample before and after potassium deposition.  Pronounced changes have been observed in both the Fermi surface and band structure. After potassium deposition, the most obvious change is the Fermi surface topology near the M points: it changes from the initial ``four intensity spots" pattern before deposition (Fig. 6a1)  into a circle after deposition (Fig. 6a2).  In the meantime, the band structure near M points exhibits dramatic changes: the initial three bands $\delta$, $\eta$ and $\varepsilon$  (Fig. 6b5 and 6b6) are replaced by an obvious electron-like shallow band after potassium deposition (Fig. 6c5 and 6c6).  There are some signatures of residual $\varepsilon$ band (Fig. 6c5) and $\delta$ band (Fig. 6c6.  We believe these bands are from the second FeSe layer that is not electron-doped by potassium deposition.  Such electronic structure evolution with potassium deposition is similar to those in MBE-grown FeSe/SrTiO$_3$ films\cite{Miyata2015,Tang2015,Song2016,Wen2016} and bulk FeSe\cite{Ye2015}, and is consistent with the transition from the N phase to the S phase found in MBE-grown FeSe/SrTiO$_3$ films\cite{He2013,LiuX2014}.  On the other hand, for the bands across $\Gamma$,  there is a slight downward shift of the hole-like $\beta$ band accompanied by a slight shrinking of the hole-like Fermi surface sheet (compare Fig. 6b4 and 6c4).  These results indicate that potassium deposition introduces electrons into the top layer of the FeSe/CaF$_2$ films.  From the area of the electron-like Fermi surface near M2, the doped electrons by potassium deposition into this sample is $\sim$0.14 electrons per Fe.

Figure 7 shows the electronic structure evolution with potassium doping for the S14K sample. Here the potassium deposition was carried out in three consecutive steps in order to follow the evolution more closely.  Overall, the evolution of the Fermi surface and band structure with potassium doping is similar to the above S4K sample. Around the $\Gamma$ point, with increasing potassium deposition, the overall band gradually shifts downwards (from 7b1, to c1, d1 and e1), accompanied by a slight shrinking of the hole-like Fermi pocket around the $\Gamma$ point (Fig. 7a1 to a2, a3 and a4).   On the other hand, around the M points, the initial three bands $\delta$, $\eta$ and $\varepsilon$ are replaced by two new bands: one is the electron-like parabola, and the other is a hole-like band that lies at a high binding energy with its top at $\sim$0.1 eV. The circular electron-like Fermi surface gets larger with increasing potassium deposition, as seen clearly from Fig. 7a1-a4, as well as from the MDCs (momentum distribution curves) along the cut2 at the Fermi level (Fig. 7f).  The resultant electron doping after the third potassium deposition is estimated to be $\sim$0.12 electrons per Fe from the area of the electron pocket near M2 (Fig. 7a4).   Again, one can see some residual bands coming from the second FeSe layer in the sample.   The overall electronic structure evolution is  consistent with the fact that potassium deposition introduces electrons, however,  we note that  it is not a rigid band shift, as seen clearly from the band evolution at M points (Fig. 7b,c,d and e).  At low temperatures,  for the FeSe films without potassium deposition, the bands at M points originate from band splitting from the nematic phase.  The introduction of electrons due to potassium deposition suppresses such a nematic transition, and eventually can totally remove the nematic transition when the electron doping is sufficiently high.  This is the main effect that can account for the dramatic change of band structure near M with potassium deposition.  The effective electron doping of the superconducting FeSe/CaF$_2$ films by potassium doping provides an opportunity to investigate the doping evolution of the superconductivity; this work will be carried out separately in the near future.

In figure 8, we compare the electronic structure of the PLD-grown multiple-layer FeSe/CaF$_2$ film with that of the MBE-grown 20-layer FeSe/SrTiO$_3$ film.  Such a comparison is intriguing because the PLD-grown multiple-layer FeSe/CaF$_2$ film is superconducting with a T$_c$ at 14 K while the MBE-grown 20-layer FeSe/SrTiO$_3$ film is non-superconducting.  Qualitatively, the electronic structure of these two kinds of films is rather similar, in terms of both the Fermi surface and band structure at  $\Gamma$ and M points. They are both in the nematic phase state at low temperature. Quantitatively, a notable difference is the hole-pocket difference around the $\Gamma$ point: the Fermi momentum of the outer $\beta$ band in PLD-grown film (Fig. 8b4, 0.23 $\pi$/a)  is much larger than that in MBE-grown film (Fig. 8c4, 0.12 $\pi$/a).  On the other hand, the bands near M points are rather similar for the two samples, e.g., the band top of the hole-like $\eta$ band at M lies at nearly the same position in PLD-grown film (Fig. 8b6, 60 meV) and in MBE-grown film (Fig. 8c6, 58 meV).
These results indicate that the PLD-grown FeSe films seem to have more holes than that in MBE-grown FeSe films.  Such a difference may come from different preparation methods that result in different sample compositions.  It is interesting to note that, sufficient electron doping can transform FeSe from N phase into S phase with high superconducting transition temperature up to 65 K\cite{He2013}. On the other hand, application of high pressure on FeSe can also enhance its T$_c$ but in this case the charge carrier is found to be hole-dominated\cite{Sun2017}.  Whether superconductivity occurs in PLD-grown multiple-layer FeSe/CaF$_2$ films is due to extra hole-doping needs further investigations.


In summary, we have carried out comprehensive ARPES investigations on the electronic structure of single crystal multiple-layer FeSe films grown on CaF$_2$ substrate by pulsed laser deposition method. Fermi surface, band structure and their temperature dependence for three kinds of samples with different superconducting transition temperatures of 4 K, 9 K and 14 K are measured.  Overall, they share similar electronic structure but sample-dependent difference is identified. These samples exhibit nematic phase transition with a nematic transition temperature between 160 K and 140 K that is higher than that of bulk FeSe.  Electron doping by potassium deposition can suppress the nematic phase transition, and transform their electronic structure from the N phase to the S phase with electron-like Fermi surface around the Brillouin zone corners.  The overall electronic structure of our PLD-grown multiple-layer FeSe films is similar to that of bulk FeSe and MBE-grown multiple-layer FeSe films.  Hole-doping difference is observed between superconducting PLD-grown multiple-layer FeSe films and non-superconducting MBE-grown multiple-layer FeSe films. These work demonstrate that the quality of the PLD-grown FeSe/CaF$_2$ films is high enough for carrying out high resolution ARPES measurements that lays a foundation for further experimental study on the superconducting gap and superconductivity mechanism.  Our ARPES results on the PLD-grown FeSe/CaF$_2$ films also establish a link between bulk FeSe single crystal and MBE-grown FeSe/SrTiO$_3$ films and will provide important information to understand the origin of high temperature superconductivity in these systems.
\vspace{3mm}

\noindent {\bf Acknowledgement} \\
Project supported by the National Natural Science Foundation of China (Grant No. 11574360), the National Basic Research Program of China (Grant Nos. 2015CB921300, 2013CB921700, and 2013CB921904), the National Key Research and Development Program of China (Grant No. 2016YFA0300300), and the Strategic Priority Research Program (B) of the Chinese Academy of Sciences (Grant No. XDB07020300).
\vspace{3mm}





\newpage

\begin{figure*}[tbp]
\begin{center}
\includegraphics[width=1.0\columnwidth,angle=0]{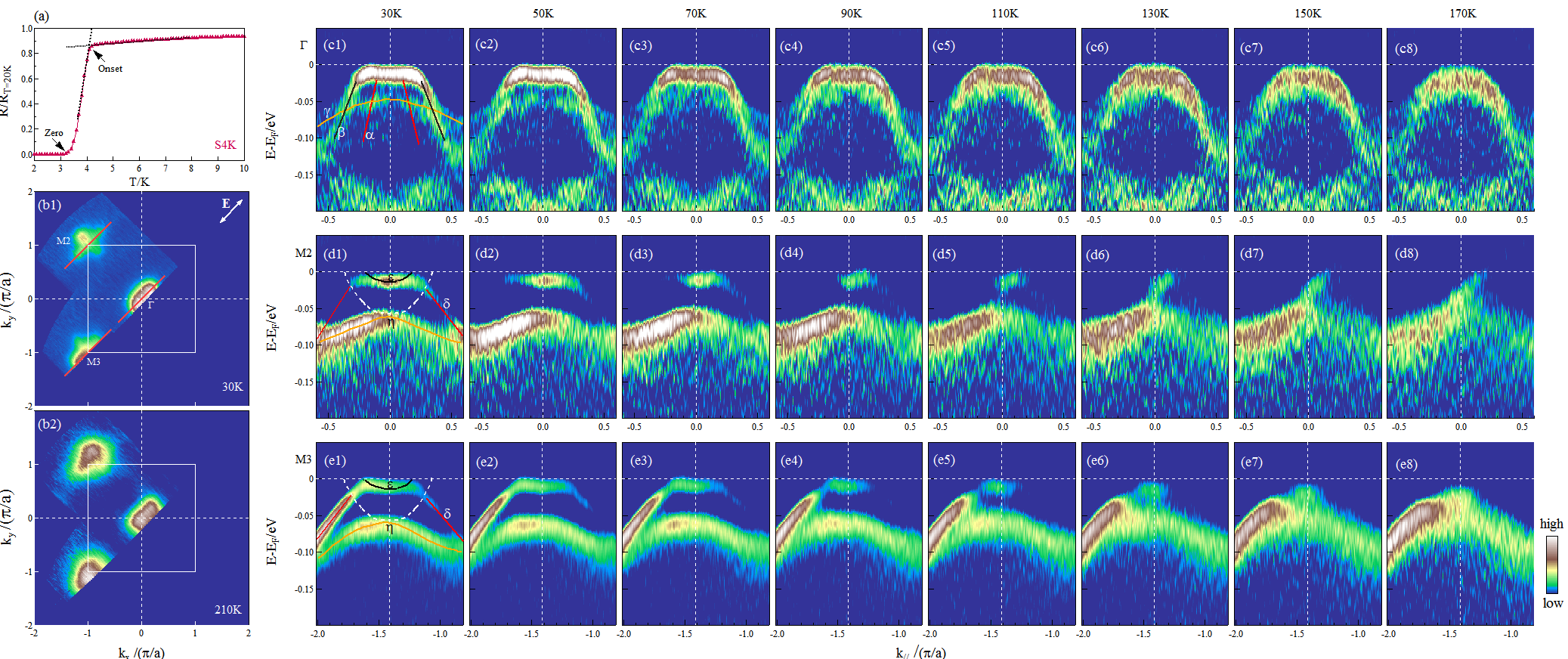}
\end{center}
\caption{Fermi surface and band structure for the FeSe/CaF$_2$  films with a T$_c$=4K. (a) Resistivity measurement of the sample. The T$_c$ onset is about 4 K.  (b1) and (b2) show Fermi surface mapping measured at a low temperature of 30 K and a high temperature of 210 K. They are obtained by integrating the spectral weight within a narrow energy window of [-20meV, 10meV] with respect to the Fermi level.  (c1) to (c8) show band structure measured at different temperatures along the momentum cut that crosses the $\Gamma$ point.  (d1) to (d8) and (e1) to (e8) show the results along the other two momentum cuts that  cross M2 and M3 respectively. The locations of the three momentum cuts are marked by red lines in (b1).   The second derivative images with respect to energy are presented here in order to  show the band structure more clearly.  The corresponding temperature is shown above these band images.  The observed bands are labeled in (c1), (d1) and (e1).
}
\end{figure*}

\begin{figure*}[tbp]
\begin{center}
\includegraphics[width=1.0\columnwidth,angle=0]{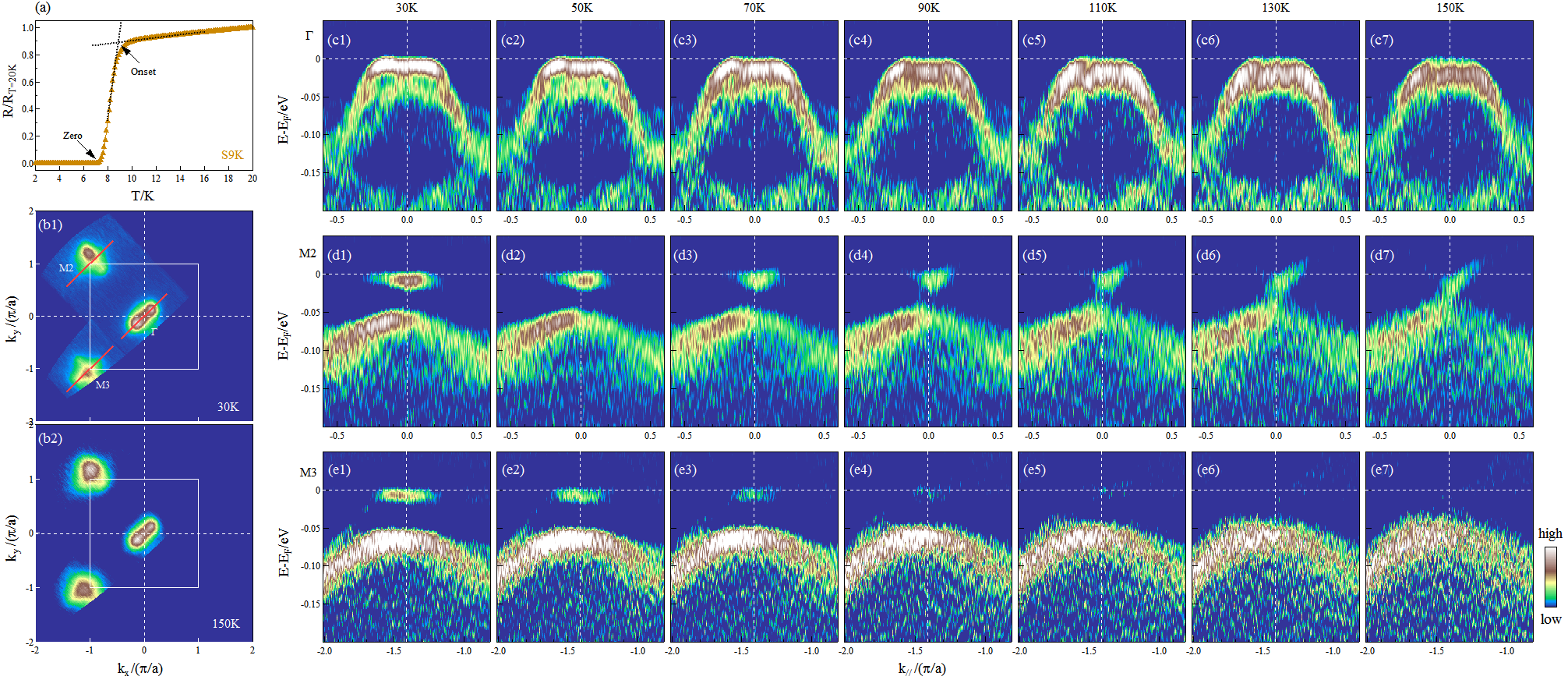}
\end{center}
\caption{Fermi surface and band structure for the FeSe/CaF$_2$  films with a T$_c$=9K. (a) Resistivity measurement of the sample. The T$_c$ onset is about 9 K.  (b1) and (b2) show Fermi surface mapping measured at a low temperature of 30 K and a high temperature of 150 K. They are obtained by integrating the spectral weight within a narrow energy window of [-20meV, 10meV] with respect to the Fermi level.  (c1) to (c7) show band structure measured at different temperatures along the momentum cut that crosses the $\Gamma$ point.  (d1) to (d7) and (e1) to (e7) show the results along the other two momentum cuts that  cross M2 and M3 respectively. The locations of the three momentum cuts are marked by red lines in (b1). The second derivative images with respect to energy are presented here in order to  show the band structure more clearly.  The corresponding temperature is shown above these band images.
}
\end{figure*}

\begin{figure*}[tbp]
\begin{center}
\includegraphics[width=1.0\columnwidth,angle=0]{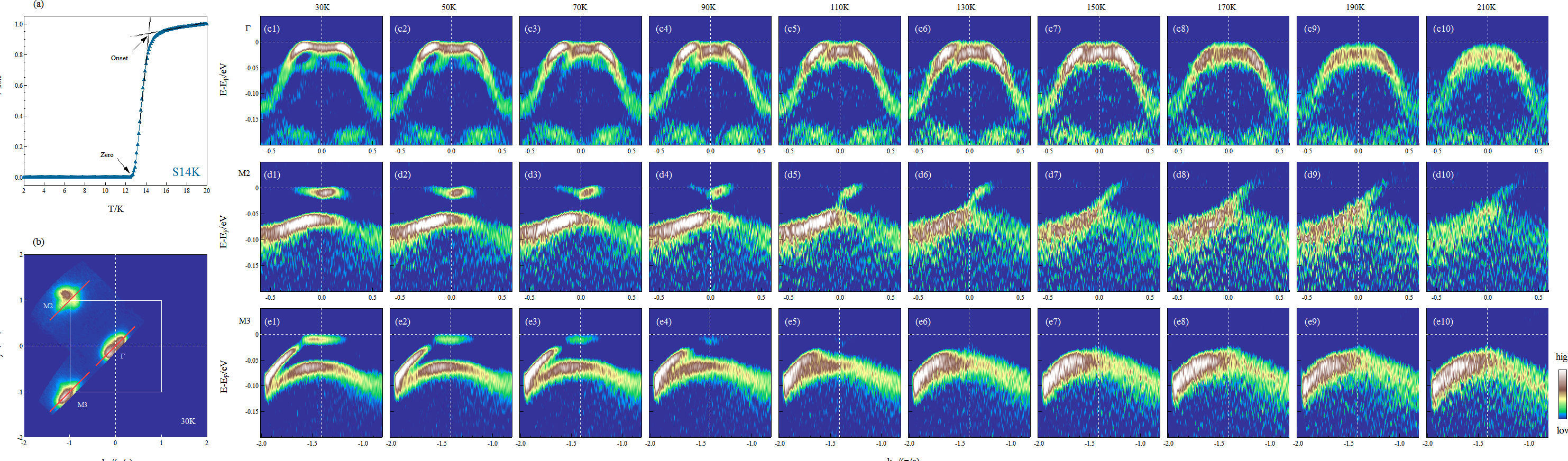}
\end{center}
\caption{Fermi surface and band structure for the FeSe/CaF$_2$  films with a T$_c$=14K. (a) Resistivity measurement of the sample. The T$_c$ onset is about 14 K.  (b)  shows Fermi surface mapping measured at a low temperature of 30 K. It is obtained by integrating the spectral weight within a narrow energy window of [-20meV, 10meV] with respect to the Fermi level.  (c1) to (c10) show band structure measured at different temperatures along the momentum cut that crosses the $\Gamma$ point.  (d1) to (d10) and (e1) to (e10) show the results along the other two momentum cuts that cross M2 and M3 respectively. The locations of the three momentum cuts are marked by red lines in (b). The second derivative images with respect to energy are presented here in order to  show the band structure more clearly.  The corresponding temperature is shown above these band images.
}
\end{figure*}

\begin{figure*}[tbp]
\begin{center}
\includegraphics[width=1.0\columnwidth,angle=0]{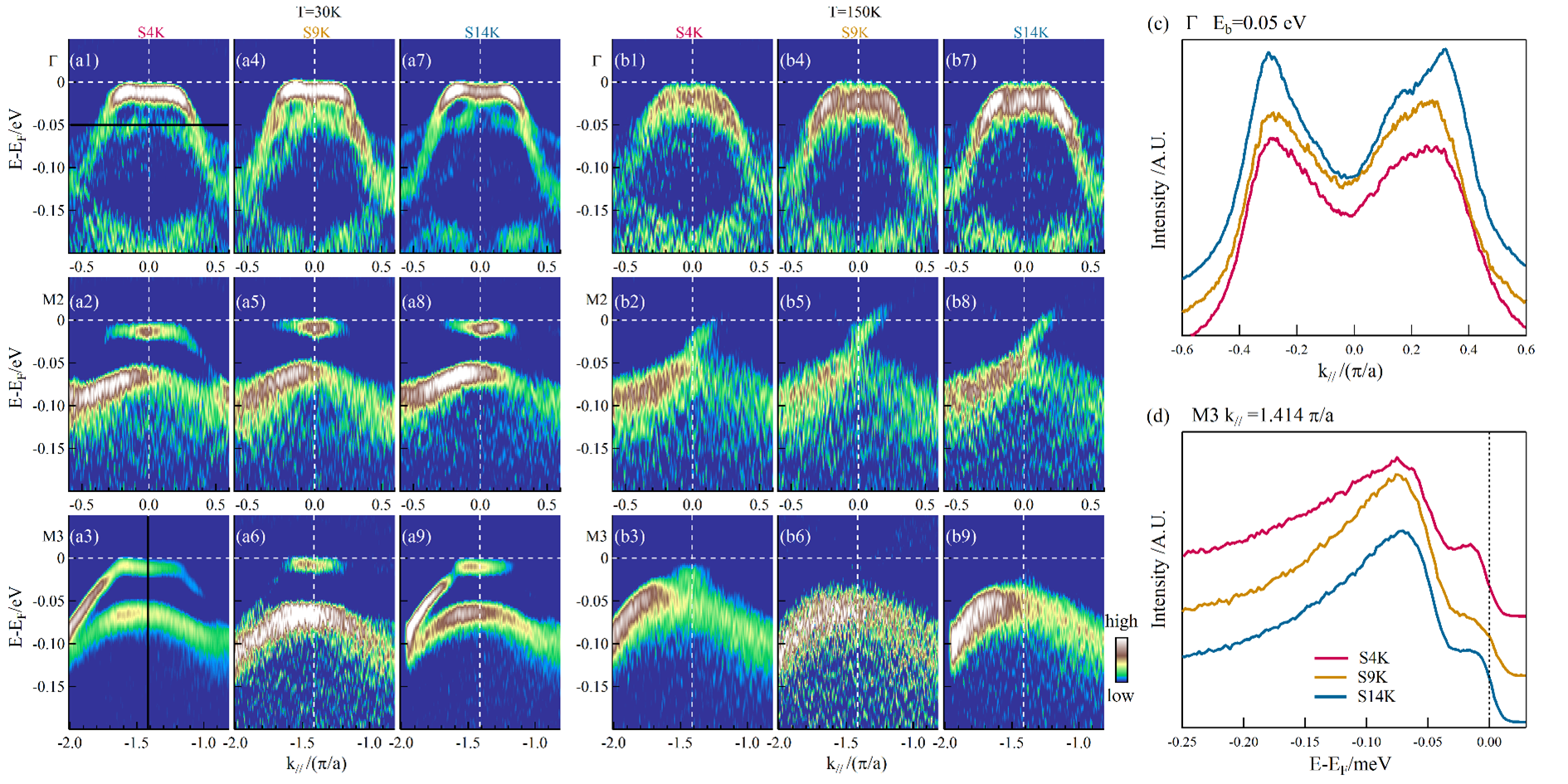}
\end{center}
\caption{Band structure comparison for the three S4K, S9K and S14K samples.  The left panel shows the data measured at 30 K while the right panel shows the data measured at 150 K.
 (a1) to (a3)  show band structure measured at 30 K for the S4K sample along momentum cuts crossing $\Gamma$, M2 and M3,  respectively. The location of the three momentum cuts is the same as that shown in Fig. 1a1. (a4)-(a6) and (a7)-(a9) show band structures along the three momentum cuts measured at 30 K for the S9K and S14K samples, respectively.   (b1)-(b3), (b4)-(b6) and (b7)-(b9) show band structure measured at 150 K along the three momentum cuts for the S4K, S9K and S14K samples, respectively.  The images are all obtained by taking second derivative with respect to energy on the original data.
}
\end{figure*}

\begin{figure*}[tbp]
\begin{center}
\includegraphics[width=1.0\columnwidth,angle=0]{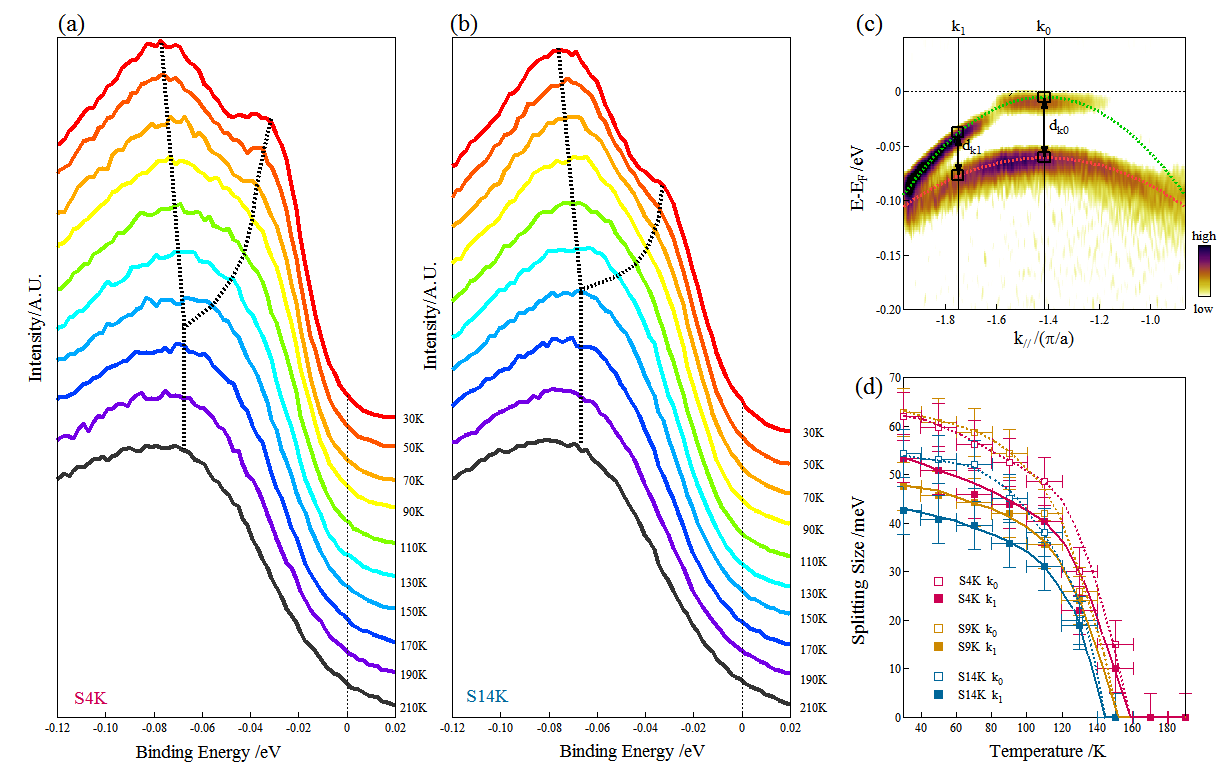}
\end{center}
\caption{Temperature dependence of the nematic phase transition in FeSe/CaF$_2$ films.  (a) and (b) show photoemission spectra (EDCs) measured at different temperatures at the momentum point k$_1$ for the S4K and S14K samples, respectively.  The dashed lines are guide to eyes for locating the peak positions.  The location of the momentum point k$_1$ is marked in (c) which shows the band structure of the S14K sample measured at 30 K along the momentum cut crossing M3 shown in Fig. 3b.   The k$_0$ in (c) represents the M point. The two bands in (c) are fitted by parabolas.  (d) shows the distance between the two hole-like bands in (c) at k$_1$ and k$_0$ points for the S4K, S9K and S14K samples.
}
\end{figure*}

\begin{figure*}[tbp]
\begin{center}
\includegraphics[width=1.0\columnwidth,angle=0]{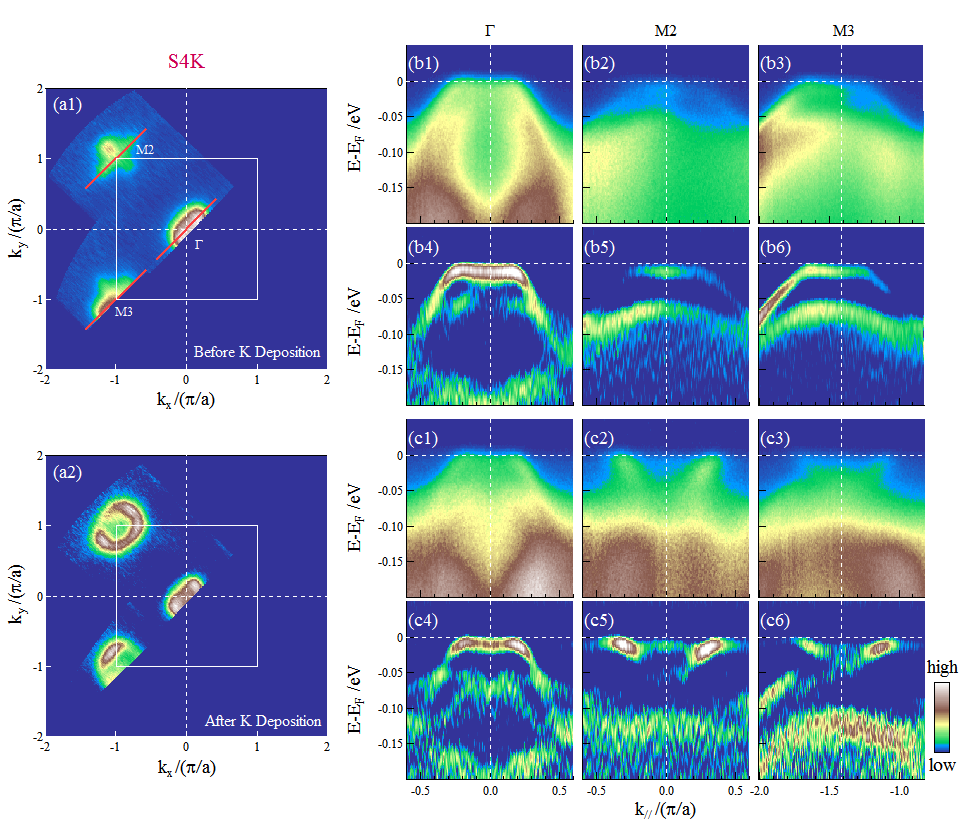}
\end{center}
\caption{Fermi surface and band structure of the S4K sample measured at 30 K before and after potassium deposition.  (a1) and (a2) are Fermi surface mapping of the S4K sample measured before and after potassium deposition, respectively.  They are obtained by integrating the spectral weight with [-20meV, 10meV] energy window with respect to the Fermi level. (b1)-(b3) show the original data measured on the S4K sample before potassium deposition along three momentum cuts crossing $\Gamma$, M2 and M3, respectively. The locations of the three momentum cuts are marked in (a1) by red lines. (b4)-(b6) are second derivative images with respect to energy corresponding to their original data (b1)-(b3).  (c1)-(c3) and (c4)-(c6) show original data along three momentum cuts after potassium deposition, and their corresponding second derivative images, respectively.
}
\end{figure*}

\begin{figure*}[tbp]
\begin{center}
\includegraphics[width=1.0\columnwidth,angle=0]{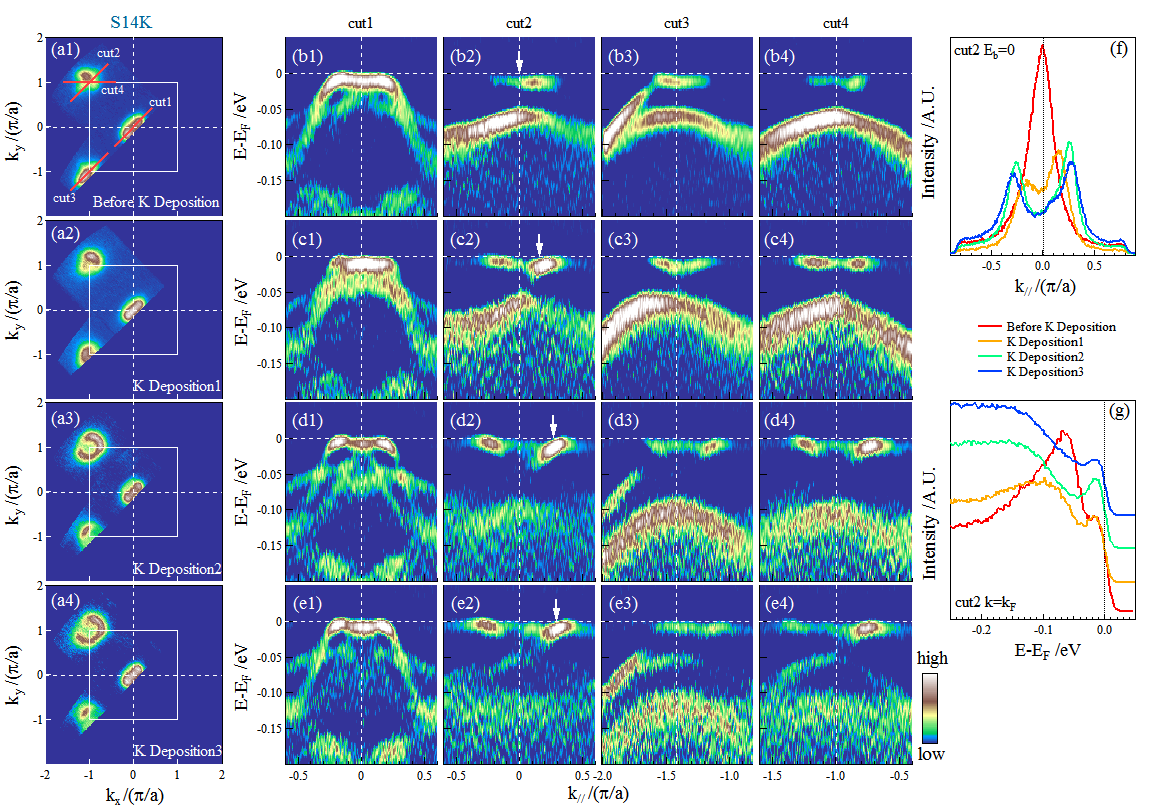}
\end{center}
\caption{Fermi surface and band structure of the S14K sample measured at 30 K before and after potassium deposition.  (a1) show Fermi surface mapping before potassium deposition, and (a2)-(a3) show Fermi surface mapping after three consecutive potassium depositions.  They are obtained by integrating the spectral weight with [-20meV, 10meV] energy window with respect to the Fermi level.  (b1)-(b4) show band structure measured for the sample before potassium deposition along four momentum cuts; their locations are marked in (a1) as red lines.     (c1)-(c4), (d1)-(d4) and (e1)-(e4) show band structure along the four momentum cuts after three consecutive potassium depositions. These images are obtained by taking second derivatives with respect to energy on the original data.   (f) shows the MDCs at the Fermi level for the cut crossing M2 point  (cut 2) before K deposition and after three K depositions.  (g) shows the EDCs at the Fermi momentum $k_F$ for the M2 cut (cut 2)  before and after three K depositions.
}
\end{figure*}

\begin{figure*}[tbp]
\begin{center}
\includegraphics[width=1.0\columnwidth,angle=0]{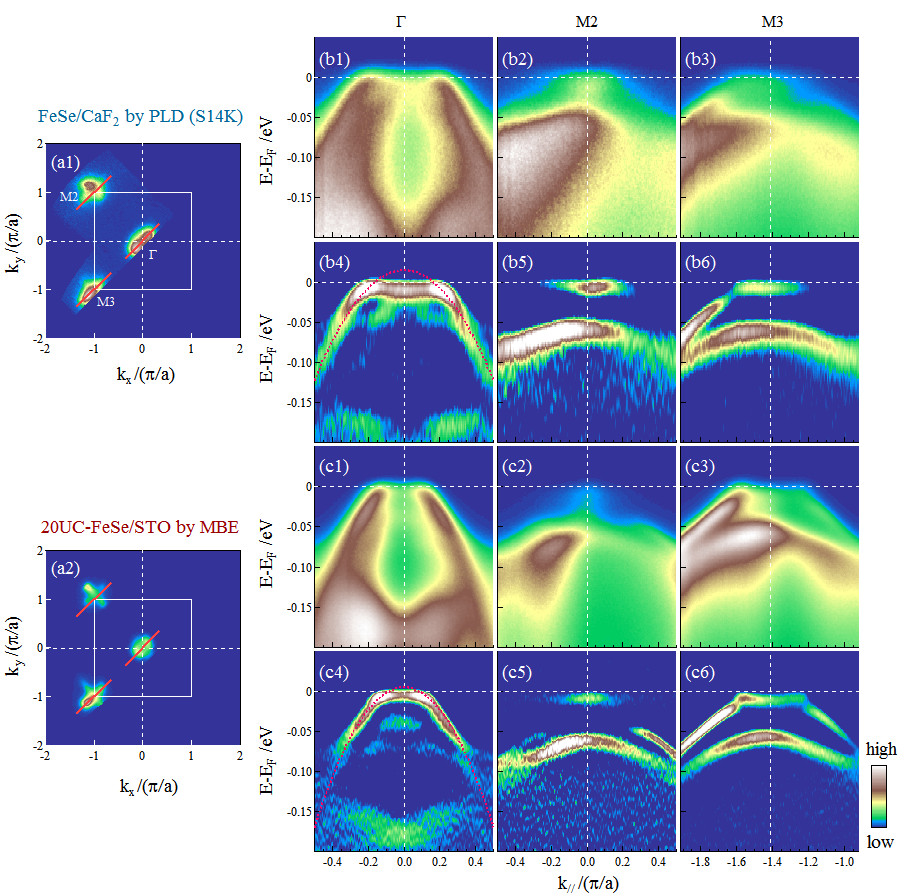}
\end{center}
\caption{Comparison of electronic structure between the PLD-grown multiple-layer FeSe/CaF$_2$ film and the MBE-grown 20-layer FeSe/SrTiO$_3$ film.   (a1) and (a2) show Fermi surface mapping of the  PLD-grown S14K sample and  MBE-grown 20-layer film, respectively, measured at 30 K.    (b1)-(b3) show original data of the band structure for PLD-grown S14K sample measured along three momentum cuts. The locations of the momentum cuts are marked in (a1) by red line.    (b4)-(b6) show corresponding second derivative images with respect to energy corresponding to (b1)-(b3). (c1)-(c3) represent original data of band structure measured along the three momentum cuts; their corresponding second derivative images are shown in (c4)-(c6).
}
\end{figure*}

\end{document}